\documentclass[11pt,epsf]{article}
\usepackage{amsmath}
\usepackage{amsfonts}
\usepackage{amssymb}
\usepackage{graphicx}
\usepackage{color}
\usepackage{comment}
\newtheorem{theorem}{Theorem}

\DeclareMathOperator*{\E}{\mathbb{E}}
\newcommand {\dfn} {\stackrel{\Delta} {=}}
\newcommand {\exe} {\stackrel{\cdot} {=}}
\newcommand {\lexe} {\stackrel{\cdot} {\le}}

\newcommand {\Rh} {R_{\mbox{\tiny h}}}

\newcommand {\reals} {{\rm I\!R}}

\newcommand {\bE} {\mbox{\boldmath $E$}}

\newcommand{\calA}{{\cal A}}

\newcommand{\calC}{{\cal C}}

\newcommand{\calE}{{\cal E}}

\newcommand{\calM}{{\cal M}}
\newcommand{\calN}{{\cal N}}

\newcommand{\calS}{{\cal S}}
\newcommand{\calT}{{\cal T}}

\newcommand{\calX}{{\cal X}}

\newcommand {\tsigma} {\tilde{\sigma}}

\allowdisplaybreaks

\begin{document}
\thispagestyle{empty}
\title{On Error Exponents of Encoder--Assisted Communication Systems
}
\author{Neri Merhav
}
\date{}
\maketitle

\begin{center}
The Andrew \& Erna Viterbi Faculty of Electrical Engineering\\
Technion - Israel Institute of Technology \\
Technion City, Haifa 32000, ISRAEL \\
E--mail: {\tt merhav@ee.technion.ac.il}\\
\end{center}
\vspace{1.5\baselineskip}
\setlength{\baselineskip}{1.5\baselineskip}

\begin{abstract}
	We consider a point--to--point communication system, where in addition to the encoder and the decoder, 
	there is a helper that observes non--causally the
	realization of the noise vector and provides a (lossy) rate--$R_{\mbox{\tiny h}}$ 
	description of it to the encoder ($R_{\mbox{\tiny h}} < \infty$). 
	While Lapidoth and Marti (2020) derived coding theorems,
	associated with achievable channel--coding rates (of the main encoder) for this model, 
	here our focus is on error exponents. We consider both continuous--alphabet, additive white Gaussian channels
	and finite--alphabet, modulo--additive channels, and for each one of them, we study the cases of both fixed--rate and 
	variable--rate noise descriptions by the helper. Our main finding is that, 
	as long as the channel--coding rate, $R$, is below the helper--rate, $R_{\mbox{\tiny h}}$, the achievable 
	error exponent is unlimited (i.e., it can be made arbitrarily large), and in some of the cases, 
	it is even strictly infinite (i.e., the error probability can be made strictly zero). However, in the range of coding rates
	$(R_{\mbox{\tiny h}},R_{\mbox{\tiny h}}+C_0)$, $C_0$ being the ordinary channel capacity (without help), the best achievable error exponent is
	finite and strictly positive, although there is a certain gap between our upper bound (converse bound) and lower bound (achievability) on
	the highest achievable error exponent. This means that the model of encoder--assisted communication is essentially equivalent to a model, where
	in addition to the noisy channel between the encoder and decoder, there is also a parallel noiseless bit--pipe of capacity $R_{\mbox{\tiny h}}$.
	We also extend the scope to the Gaussian multiple access channel (MAC) and characterize the rate sub--region, where the achievable error exponent is
	unlimited or even infinite.\\

\noindent
	{\bf Index Terms:} error exponent, encoder--assisted, sphere--packing, multiple--access channel, additive channel.
\end{abstract}

\clearpage
\section{Introduction}

In a recent work \cite{LM20}, Lapidoth and Marti (see also Marti \cite{Marti19}) have studied the problem of coded communication, where 
in addition to the usual encoder and decoder, there is also a helper that observes
(causally or non--causally) the realization of the channel--noise vector and provides 
the encoder with a description of this noise vector at the rate of $R_{\mbox{\tiny h}}$ bits (or nats) per
noise sample on the average.\footnote{I.e., the description is of total length of $nR_{\mbox{\tiny h}}$ for a noise vector of length $n$, which is
the block length of the channel encoder.}
The underlying motivation, described in \cite{LM20} for the non--causal case, is a scenario where the encoder is located in the vicinity of
an interfering transmitter (in the role of the helper), 
the `noise', in this context, is the codeword that this interferer is about to transmit, and the interferer is connected to
the main encoder by a rate--limited, noiseless bit--pipe. Lapidoth and Marti have provided, in that work, 
coding theorems, that characterize the capacity of such a system.\footnote{Note that in the limit of $R_{\mbox{\tiny h}}\to\infty$, 
this problem setup becomes a degenerated special case of the Gel'fand--Pinsker model \cite{GP80},
and in particular, the dirty paper channel model \cite{Costa83}, when there is no additional noise vector that is not known to the transmitter.}
For the case where the channel is Gaussian (and a few other channels), Lapidoth and Marti  
have proved that this capacity is given by $C_0+R_{\mbox{\tiny h}}$, where $C_0$ is the ordinary capacity
of the Gaussian channel (without help). It is interesting to point out 
that their capacity--achieving coding scheme is based on the notion of {\it flash--help},
which means allocating the entire helper rate budget to a very accurate description of an extremely small part of 
the noise vector (and leaving no remaining helper bits for the other part of the noise vector), 
rather than spreading rate budget uniformly across the entire noise vector of length $n$.

In this work, we study the problem of encoder--assisted communication from the aspect of achievable error exponents. 
As in \cite{LM20}, we also consider coding schemes that are based on the idea of flash--help, but our schemes are 
somewhat different from the one in \cite{LM20}, as we are in the quest of the more refined objective of maximizing the error exponent at a given rate,
rather than maximizing the achievable rate. We consider both fixed--rate and variable--rate lossy compression by the helper, 
and both the continuous--alphabet,
additive Gaussian noise (AWGN) channel, and the finite--alphabet, modulo--additive channel.
We show that as long as the coding rate, $R$, of the main encoder is less than the helper rate, $R_{\mbox{\tiny h}}$, the error exponent that can be achieved
is arbitrarily large, and in some of the cases considered, it is even strictly infinite, as the error probability may vanish to zero.\footnote{While this
result would not have seemed surprising had the error exponent been defined with respect to the small segment in which the noise is accuraltey described
(and hence could be essentially canceled by the transmitter, 
in the flash--help approach), it is not quite trivial that it is still true even when the error exponent is
defined with respect to the entire block length, $n$, as usual.}
For the additive white Gaussian channel, in the range of rates, $R_{\mbox{\tiny h}} < R < R_{\mbox{\tiny h}}+C_0$, 
our coding scheme achieves a finite error exponent, given by 
$E_{\mbox{\tiny a}}(R-R_{\mbox{\tiny h}})$, where
$E_{\mbox{\tiny a}}(\cdot)$ is any achievable error exponent of 
ordinary channel coding, without help (e.g., the random coding exponent, the expurgated exponent, etc.). Thus, the achieved error exponent is positive
for any rate below $R_{\mbox{\tiny h}}+C_0$, as expected based on \cite{LM20}. 

We also derive an upper bound 
(converse bound) on the maximum achievable error exponent, which is a weakened\footnote{Deriving a converse bound for this model is a non--trivial task, as
the transmitted signal and the noise are correlated in an arbitrary manner.}
version of the sphere--packing bound, henceforth referred to as the {\it weak sphere--packing} (WSP) bound. 
While the WSP bound may not be tight in the usual quantitative sense, we believe that it is at least valuable in the sense
of matching the achievability results in the qualitative sense, as its value is infinite for $R < R_{\mbox{\tiny h}}$, finite but positive for
$R_{\mbox{\tiny h}} < R < R_{\mbox{\tiny h}}+C_0$, and zero for $R\ge R_{\mbox{\tiny h}}+C_0$. This means that these three different types of behavior of the
error exponent function are inherent to the model being addressed, and not only a property of the specific coding scheme we propose. It also indicates that
in a certain sense, this system configuration is equivalent to the existence of an additional, parallel noiseless bit--pipe of capacity $R_{\mbox{\tiny h}}$
between the encoder and decoder. As long as $R < R_{\mbox{\tiny h}}$, perfectly reliable 
transmission takes place solely via the noiseless bit--pipe. When $R$ exceeds
$R_{\mbox{\tiny h}}$, the excess rate, $R-R_{\mbox{\tiny h}}$, is transmitted via the original channel, without help, and the 
resulting error exponent is $E_{\mbox{\tiny a}}(R-R_{\mbox{\tiny h}})$.

Finally, as in \cite{LM20}, we also outline a few modifications and extensions of the scope to:
(i) general continuous--alphabet, memoryless additive channels, (ii) modulo--additive channels, and
(iii) the Gaussian multiple access channel (MAC), where in the latter,
help is provided to both encoders, and the total help rate of $R_{\mbox{\tiny h}}$, is optimally divided between the two encoders.
Here too, there are three different regions
in the plane of rates: the region of infinite error exponent, the region of finite error exponent, and the region of zero error exponent, which is
the complement of the capacity region of the Gaussian MAC.

The outline of this paper is as follows. In Section \ref{nc}, we establish the notation conventions. In Section \ref{pso}, we formulate the problem and
spell out the objectives. In Section \ref{mr}, we provide the main results for the AWGN channel and discuss them.
In Section \ref{ma}, we outline the parallel derivations and results for the modulo--additive channel, and finally, in Section \ref{mac},
we do the same for the Gaussian MAC.

\section{Notation Conventions}
\label{nc}

Throughout the paper, random variables will be denoted by capital
letters, specific values they may take will be denoted by the
corresponding lower case letters, and their alphabets
will be denoted by calligraphic letters. Random
vectors and their realizations will be denoted,
respectively, by capital letters and the corresponding lower case letters,
superscripted by their dimensions. Their alphabets will also be superscripted by their
dimensions. For example, the random vector $X^n=(X_1,\ldots,X_n)$, ($n$ --
positive integer) may take a specific vector value $x^n=(x_1,\ldots,x_n)$
in $\calX^n$, the $n$--th order Cartesian power of $\calX$, which is
the alphabet of each component of this vector.
Sources and channels will be denoted by capital letters,
subscripted by the names of the relevant random variables/vectors and their
conditionings, if applicable, following the standard notation conventions,
e.g., $Q_X$, $P_{Y|X}$, and so on. When there is no room for ambiguity, these
subscripts will be omitted.
The probability of an event $\calE$ will be denoted by $\mbox{Pr}\{\calE\}$,
and the expectation
operator with respect to (w.r.t.) a probability distribution $P$ will be
denoted by
$\E_P\{\cdot\}$. Again, the subscript will be omitted if the underlying
probability distribution is clear from the context.
The entropy of a generic random variable $X$, with a distribution $Q$ on $\calX$, will be denoted by
$H_Q(X)$. 
The Kullback--Leibler divergence between two probability distributions,
$Q$ and $P$ with a common alphabet, say, $\calX$, is defined as
\begin{equation}
        D(Q\|P)=\sum_{x\in\calX}Q(x)\log\frac{Q(x)}{P(x)},
\end{equation}
where logarithms, here and throughout the sequel, are understood to be taken to the base $\mbox{e}$, unless specified otherwise.
Similarly, the divergence between two pdfs will be defined in the same manner except that the summation will be replaced by an integral.

For two
positive sequences $a_n$ and $b_n$, the notation $a_n\exe b_n$ will
stand for equality in the exponential scale, that is,
$\lim_{n\to\infty}\frac{1}{n}\log \frac{a_n}{b_n}=0$. Similarly,
$a_n\lexe b_n$ means that
$\limsup_{n\to\infty}\frac{1}{n}\log \frac{a_n}{b_n}\le 0$, and so on.
The cardinality of a finite set, $\calA$, will be denoted by $|\calA|$.

\section{Problem Setting and Objectives}
\label{pso}

Consider a memoryless additive channel,
\begin{equation}
	Y_i=X_i+Z_i,~~~~~~~~i=1,2,\ldots,n
\end{equation}
where $X_i$ is a real valued random variable (RV), designating the channel input at time $i$ and $\{Z_i\}$ are independently and identically
distributed (i.i.d.) RV's, whose probability density function (PDF) is denoted by $f(z)$. The sequence $Z^n=(Z_1,\ldots,Z_n)$ designates the noise
vector. Finally, $Y_i$ is the channel output at time $i$.

The system configuration considered is the same as in \cite{LM20}. It consists of three users: a transmitter (encoder), a receiver (decoder) and a helper.
The helper observes the realization $z^n$ of the noise vector and transmits (non--causally, in general) to the encoder
a description of this vector using $nR_{\mbox{\tiny h}}$ nats via a noiseless link. In mathematical terms, the helper is defined by 
a function $T:\reals^n\to\calT=\{0,1,\ldots, e^{nR_{\mbox{\tiny h}}}-1\}$. The encoder receives the helper's message, $T(z^n)$, as well as
an ordinary information message index, $m\in\calM=\{0,1,\ldots,e^{nR}-1\}$,
$R$ being the coding rate in nats per channel use, and generates a channel input vector, $x^n=\phi(m,T(z^n))$, where $\phi:\calM\times\calT\to\calC\subseteq
\reals^n$. The message index, $m$, is assumed a RV, uniformly distributed across $\calM$, and independent of $Z^n$.
The channel input vector must obey a generalized power constraint,
\begin{equation}
	\sum_{i=1}^n\E\{\rho([\phi(m,T(Z^n))]_i)\}\le nP,
\end{equation}
where $P$ is the allowed generalized power level and the expectation is taken with respect to (w.r.t.) the randomness of both $Z^n$ and $m$.
Here, $\rho:\calX\to\reals^+$ is the {\it generalized power function} and $[\phi(m,T(Z^n))]_i$ designates the $i$--th component of the codeword,
$x^n=\phi(m,T(z^n))$, $i=1,2,\ldots,n$.
Finally, the decoder is defined by a mapping $\psi:\reals^n\to\calM$, and $\hat{m}=\psi(y^n)$ denotes the decoded message.
The {\it probability of error} is defined as
\begin{equation}
	P_{\mbox{\tiny e}}(\phi,\psi,T)=\mbox{Pr}\{\psi(\phi(m,T(Z^n))+Z^n)\ne m\}.
\end{equation}
For a given $R_{\mbox{\tiny h}}$, an {\it achievable rate} is a coding rate, $R$, with the following property: for every $\epsilon > 0$,
there exists a sufficiently large block length, $n$, such that there exist a 
rate--$R$ encoder, $\phi$, a decoder, $\psi$, and a helper, $T$,
so that $P_{\mbox{\tiny e}}(\phi,\psi,T)\le\epsilon$. The {\it capacity} of the system, 
$C(R_{\mbox{\tiny h}})$, is defined as the supremum of all achievable rates. 

For the case of the additive Gaussian channel, i.e., the case where
$Z_i\sim\calN(0,\sigma^2)$ for all $i$ and a quadratic power function, $\rho(x)=x^2$ (as well as a few other cases), 
Lapidoth and Marti \cite{LM20} have proved a coding theorem and its
converse theorem, which together establish the fact that
\begin{equation}
	C(R_{\mbox{\tiny h}})=R_{\mbox{\tiny h}}+C_0,
\end{equation}
where $C_0$ is the ordinary capacity of the same channel without help, which in the Gaussian case, amounts to
\begin{equation}
	C_0=c(\gamma)\dfn\frac{1}{2}\log(1+\gamma),
\end{equation}
where $\gamma\dfn P/\sigma^2$.

Our objective in this paper is to study achievable error exponents for encoder--assisted communication systems, that is, to obtain
upper and lower bounds on the {\it reliability function},
\begin{equation}
	E(R)\dfn\limsup_{n\to\infty}\left\{-\frac{\log[\inf_{\phi,\psi,T}P_{\mbox{\tiny e}}(\phi,\psi,T)]}{n}\right\},
\end{equation}
where the infimum over $\phi$ and $\psi$ is understood to be taken over all rate--$R$ encoders and their corresponding decoders.

\section{The Single--User AWGN Channel}
\label{mr}

\subsection{Achievability}

To fix ideas, we begin with the case of the additive white Gaussian noise (AWGN) channel, that is,
\begin{equation}
	f(z)=\frac{e^{-z^2/(2\sigma^2)}}{\sqrt{2\pi\sigma^2}}
\end{equation}
and $\rho(x)=x^2$, and later on we discuss a possible
extension to more general continuous--alphabet, memoryless channels and generalized power functions.
Our first result applies to fixed--rate lossy compression by the helper.

\begin{theorem}
	\label{thm1}
	Consider the setting defined in Section \ref{pso}, for the AWGN channel. Then,
	\begin{equation}
		\label{lb}
		E(R)\ge\left\{\begin{array}{ll}
			\infty & R < \Rh\\
			E_{\mbox{\tiny a}}(R-\Rh) & \Rh < R < \Rh+C_0\\
		0 & R\ge\Rh+C_0\end{array}\right.
	\end{equation}
	where:
	\begin{enumerate}
		\item the assertion $E(R)\ge\infty$, which is equivalent to $E(R)=\infty$, in the first line of (\ref{lb}), 
			should be understood in the sense that an arbitrarily
			error exponent is achievable,
		\item $E_{\mbox{\tiny a}}(\cdot)$ is any achievable error exponent function associated with AWGN channel without help, and
		\item $C_0$ is the capacity of the AWGN channel without help, that is,
		$C_0=c(\gamma)$.
	\end{enumerate}
\end{theorem}

The error exponent function, $E_{\mbox{\tiny a}}(\cdot)$, can be chosen to be the random coding exponent or the expurgated exponent of the
AWGN channel (in particular, 
the larger between the two, for the given $R$), whose expressions can be found, for example, in \cite[Subsection 7.4]{Gallager68}, 
or the error exponent associated with an arbitrary signal constellation that complies
with the power constraint.

\noindent
{\it Proof.}
Consider the following coding scheme, which is in the spirit of that of \cite{LM20}, but with a few twists.
Let us divide the block of $n$ transmitted symbols into two segments.
The first segment, of length $t=n\tau$ (for some $0<\tau< 1$), will be the segment where the 
encoder receives from the helper $nR_{\mbox{\tiny h}}$ nats of description of the corresponding segment of the noise vector, 
$z^{t}=(z_1,\ldots,z_{t})$, whereas over the remaining part of the block, of length $n-t=n(1-\tau)$, 
no help is provided at all.
A uniform scalar quantizer is used to represent each coordinate of $z^{t}$ using 
$\frac{nR_{\mbox{\tiny h}}}{t}=
\frac{R_{\mbox{\tiny h}}}{\tau}$ nats per sample. 
If $\tau$ is small, then $R_{\mbox{\tiny h}}/\tau$ is large, and the quantizer
operates in the high--resolution regime (see, e.g., \cite[Chap.\ 5]{Gray90}). More precisely, consider the $t$--dimensional hyper-sphere of radius
$\sqrt{t\sigma^2(1+s)}$, centered at the origin, in the space of noise vectors, $\{z^t\}$, where $s > 0$ 
is a design parameter, to be chosen later. The helper's lossy compression scheme is based on partitioning this hyper-sphere into hyper-cubes of size
$\Delta > 0$ and quantizing $z^t$ into the center of the hyper-cube 
to which it belongs. If $z^t$ falls outside the hyper-sphere, then the compression fails
and an error occurs. Accordingly,
the step--size, $\Delta$, of the uniform scalar quantizer is chosen such that
\begin{equation}
	nR_{\mbox{\tiny h}}=\log\left(\frac{\mbox{Vol}\{\mbox{hyper-sphere of radius $\sqrt{t\sigma^2(1+s)}$}\}}
	{\Delta^{t}}\right)\approx\frac{n\tau}{2}\log\frac{2\pi e\sigma^2(1+s)}{\Delta^2},
\end{equation}
where the second, approximate equality can be found, for example, in \cite[p.\ 144, eq.\ (7.30)]{Zamir14}, and the approximation is in the sense
that the two expressions differ by a quantity that grows sub-linearly with $n$, which will henceforth be ignored.
Equivalently,
\begin{equation}
	\Delta=\sqrt{2\pi e\sigma^2(1+s)}\cdot e^{-R_{\mbox{\tiny h}}/\tau}.
\end{equation}
Let $q(z^{t})$ be the quantized version of $z^{t}$ using this quantizer. Let the
main encoder's transmission, at this segment, be given by
\begin{equation}
	x^{t}(m,T(z^t))=\tilde{x}^{t}(m)-q(z^{t}),
\end{equation}
where $\tilde{x}^{t}(m)$ is a codeword of length $t=n\tau$ that depends only on (part of) the message.
The corresponding segment of the received signal is then
\begin{equation}
	y^{t}=x^{t}(m,T(z^t))+z^{t}=\tilde{x}^{t}(m)-q(z^{t})
	+z^{t}\dfn\tilde{x}^{t}(m)+\tilde{z}^{t},
\end{equation}
where $\tilde{z}^{t}=z^{t}-q(z^{t})$ is the residual quantization noise.

As long as the norm of $z^{t}$ is less than $\sqrt{t\sigma_z^2(1+s)}$, the quantization error vector,
$\tilde{z}^{t}$, lies within the hyper-cube $[-\frac{\Delta}{2},+\frac{\Delta}{2}]^{t}$.
Therefore, if main channel--encoder uses a simple lattice code that is 
based on a Cartesian grid of step--size $\Delta$ in each coordinate, the transmission
in this segment will be error--free, as the residual noise vector cannot cause a passage to the hyper--cube of any other codeword.
Such a lattice code can therefore support an error--free transmission of $nR'$ informatiom nats, where
\begin{eqnarray}
	nR'&=&\log\left(\frac{\mbox{Vol}\{\mbox{hyper-sphere of radius $\sqrt{tP}$}\}}
	{\Delta^{t}}\right)\nonumber\\
	&\approx&\frac{n\tau}{2}\log\frac{2\pi e P}{\Delta^2}\nonumber\\
	&=&\frac{n\tau}{2}\log\frac{2\pi e P}{2\pi e\sigma^2(1+s)e^{-2R_{\mbox{\tiny h}}/\tau}}\nonumber\\
	&=&nR_{\mbox{\tiny h}}+\frac{n\tau}{2}\log\frac{P}{\sigma^2(1+s)}.
\end{eqnarray}
Clearly, if $\tau$ tends to zero (after the limit $n\to\infty$ has been taken), then $R'$ approaches $R_{\mbox{\tiny h}}$. In other words, we can transmit
essentially $nR_{\mbox{\tiny h}}$ nats per channel use error--free, provided that $\|z^t\|^2\le t\sigma^2(1+s)$.
An error will occur, in this segment, only if $\|z^t\|^2> t\sigma^2(1+s)$.
For the case of the Gaussian channel, this probability is easily upper bounded by the Chernoff bound, which yields (see, e.g., 
\cite[Proposition 13.1.3, p.\ 374]{Zamir14}), 
\begin{equation}
	\mbox{Pr}\left\{\sum_{i=1}^{t}Z_i^2 > t\sigma^2(1+s)\right\}\le
	\exp\left\{-\frac{t}{2}[s-\ln(1+s)]\right\}=
	\exp\left\{-n\cdot\frac{\tau}{2}[s-\ln(1+s)]\right\}.
\end{equation}
For a given (arbitrarily small, but positive) $\tau$, let $s$ be sufficiently large such that
$\tau[s-\ln(1+s)]/2$ is as large as desired, say, $s=B/\tau$, where $B> 0$ is an arbitrarily large constant.
Then, the error exponent is essentially as large as $\frac{B}{2}\left[1-\frac{\tau}{B}\ln\left(1+\frac{B}{\tau}\right)\right]$, 
which for large enough $B/\tau$, is at least as large $\frac{B}{3}$, as $\lim_{u\to\infty}\frac{\ln(1+u)}{u}=0$. The number of information nats
we can encode in the short segment of length $t=n\tau$, is therefore
\begin{equation}
	nR'\approx nR_{\mbox{\tiny h}}+\frac{n\tau}{2}\log\frac{P}{\sigma^2(1+s)}=
	 nR_{\mbox{\tiny h}}+\frac{n\tau}{2}\log\frac{P\tau}{\sigma^2(B+\tau)},
\end{equation}
which is still arbitrarily close $nR_{\mbox{\tiny h}}$ for sufficiently small $\tau$.
Thus, we can transmit about $nR_{\mbox{\tiny h}}$ nats with an error exponent that is as large as desired, using
a very simple encoder. 

If $R > R_{\mbox{\tiny h}}$, we also use the second segment, of length $n(1-\tau)$, to transmit the
remaining $\Delta R\dfn R-R_{\mbox{\tiny h}}-\frac{\tau}{2}\log\frac{P\tau}{\sigma^2(B+\tau)}$
nats using an ordinary code, say, an orthogonal code, or a random 
code, or an expurgated code, without help, whose error exponent function is denoted generically by $E_{\mbox{\tiny a}}(\cdot)$. 
The error exponent at the second
segment is therefore
\begin{equation}
(1-\tau)E_{\mbox{\tiny a}}\left(\frac{\Delta R}
{1-\tau}\right).
\end{equation}
It follows that the overall error exponent is given by
\begin{equation}
\min\left\{\frac{\tau[s-\ln(1+s)]}{2}, (1-\tau)E_{\mbox{\tiny a}}\left(\frac{\Delta R}
{1-\tau}\right)\right\}.
\end{equation}
Since the first term can be made as large as desired and $\Delta R$ can be made arbitrarily close to $R-\Rh$, by
selecting $\tau$ small enough and $s$ large enough, this error exponent is dominated by the second term,
which becomes $E_{\mbox{\tiny a}}(R-R_{\mbox{\tiny h}})$. This completes the proof of Theorem \ref{thm1}. $\Box$\\

\noindent
{\bf Discussion.}
A few comments are in order:\\

\noindent
{\it The phase transition at $R=\Rh$.}
As we can see, the behavior of the error exponent function has two dichotomies: it is
essentially infinite for $R<  R_{\mbox{\tiny h}}$, but finite
in the range $R_{\mbox{\tiny h}}< R <  C_0+R_{\mbox{\tiny h}}$.
This behavior is not just a result of the particular coding 
scheme presented, it appears also in the converse bound to be presented in the sequel.
Any gap between the converse bound and the achievability bound will only concern the
exact error exponent at $R>  R_{\mbox{\tiny h}}$, but not these dichotomies from the two sides of
$R=R_{\mbox{\tiny h}}$. In other words, there is an inherent phase transition in the error exponent at
$R=\Rh$.\\

\noindent
{\it Equivalence to a noiseless bit pipe.}
Since the residual noise (i.e., the quantization error of the noise) can be made very small, thanks to the
high--resolution regime, it may not be surprising that an arbitrarily 
large error exponent can be obtained for $R < R_{\mbox{\tiny h}}$ 
if the (effective) block length is considered to be $t=n\tau$ (the length
of the ``with-help'' segment). It is 
less trivial, however, that an arbitrary large error exponent is
still achievable also when the error probability is viewed as a function of $n$. This is a considerable difference in view of the fact that $\tau$
is chosen very small. This means that the encoder--assisted coding configuration under discussion is essentially 
equivalent to a system with an additional, 
parallel noiseless bit pipe, between the transmitter and receiver, 
whose capacity is $R_{\mbox{\tiny h}}$.\\

\noindent
{\it Simplicity of implementation of the ``with--help'' phase.}
Note that the encoding in the ``with--help'' part of the coding scheme is extremely simple to implement, both at the encoder
and the decoder. The helper simply applies a uniform scalar quantizer (after verifying that $z^t$ falls within the designated hyper-sphere), the encoder 
implements a one--dimensional (Cartesian) lattice code 
(with a power limitation), and the decoder simply quantizes the channel output, again, using a scalar quantizer.\\

\noindent
{\it Variable--rate coding by the helper.}
If the helper is allowed to use a variable--rate code, it can describe $z^t$ using
$L(z^{t})\approx-\log[f(z^{t})\cdot\Delta^{t}]$ nats over the entire space,
without the limitation to a sphere with a finite radius. In this case,
the average description length is approximated by \cite[p.\ 125, eq.\ (5.4.2)]{Gray90},
\begin{equation}
nR_{\mbox{\tiny h}}\approx h(Z^{t})-t\log\Delta=
\frac{n\tau}{2}\log\frac{2\pi e\sigma^2}{\Delta^2}.
\end{equation}
Here the ``with--help'' phase is strictly
error--free, which means a strictly infinite error exponent in that phase, unless there is a finite buffer
for the noise description, and then buffer overflow yields a decoding error 
(see the parallel derivation in Subsection \ref{var} below). Also, there is no 
need for the parameter $s$ that caused the rate reduction of the fixed--rate help.\\

\noindent
{\it More general continuous--alphabet, memoryless, additive channels.}
While our proposed achievability scheme (described above in the proof of Theorem \ref{thm1}), was constructed
with the AWGN channel in mind, it is conceptually possible to extend it, under certain regularity
conditions, to an arbitrary continuous--alphabet, additive, memoryless
channel with a generalized power function, $\rho$, 
where the channel noise is still i.i.d., but with a general pdf, $f$. We will not carry out the derivation in full detail here, but only outline the
required modifications very briefly.
The hyper--sphere of noise vectors, within which the helper quantizes the noise $t$-vectors (still using a uniform scalar quantizer),
is best replaced by the set of vectors of the form, $\calS=\left\{z^t:~-\sum_{i=1}^t\log f(z_i)\le tA\right\}$ ($A$ being larger than the
differential entropy of $Z_1$), because by the Neymann--Pearson theorem, it yields the best possible trade--off between
a small volume of $\calS$ and a small probability of $\calS^{\mbox{\tiny c}}$. The channel code pertaining to the ``with--help'' segment of length $t$
is still a simple Cartesian, lattice code, but now it is confined to the generalized hyper--sphere $\left\{x^t:~\sum_{i=1}^t\rho(x_t)\le tP\right\}$.
In spite of these modifications, the basic property remains unchanged: the first $n\Rh$ information
nats can be conveyed, essentially error--free, within the ``with--help'' part, whereas the remaining nats (if any) can be encoded using an ordinary
channel code across the complementary segment, without help.

\subsection{Converse Bound}

Returning the AWGN channel with a quadratic power function,
we next provide a converse bound (i.e., an upper bound to the error exponent), 
which is a certain version of the sphere--packing bound. The fact that the
transmitted signal and the noise are allowed to be correlated (and in an arbitrary manner) causes considerable complications, and
it is not apparent how to apply the ordinary techniques of proving the sphere--packing
bound. A more general argument that bypasses this difficulty is therefore needed.
As customary with sphere--packing bounds, it is based on a change--of--measures
argument, but to avoid the complication associated with an arbitrary noise--dependent transmitter, 
this change--of measures is applied solely to the noise density, keeping the channel purely additive.
It replaces the underlying noise density, $f=\calN(0,\sigma^2)$, by $g=\calN(0,\tsigma^2)$, 
where $\tsigma^2$ is chosen such that the given rate, $R$, becomes just above
capacity. The resulting bound is
a somewhat weaker version of the sphere--packing bound, 
henceforth referred to as the {\it weak sphere--packing bound} (WSP bound),
which is given by
\begin{eqnarray}
	\label{wsp}
	E_{\mbox{\tiny wsp}}(R)&=&\left\{\begin{array}{ll}
		\infty & R < \Rh\\
	\frac{1}{2}\left[\frac{e^{2C_0}-1}{e^{2(R-R_{\mbox{\tiny h}})}-1}
		-\ln\left(\frac{2^{2C_0}-1}{e^{2(R-R_{\mbox{\tiny h}})}-1}\right)-1\right] & \Rh < R < \Rh+C_0\\
	0 & R\ge \Rh+C_0\end{array}\right.
\end{eqnarray}
Our converse--bound result is asserted as follows.
\begin{theorem}
	\label{thm2}
	For the problem setting described in Section \ref{pso}, 
	\begin{equation}
		E(R)\le  E_{\mbox{\tiny wsp}}(R).
	\end{equation}
\end{theorem}
The WSP bound shares two important properties of the achievability bound: (i) infinite error exponent for $R<  R_{\mbox{\tiny h}}$
and a finite error exponent for $R_{\mbox{\tiny h}} < R <  R_{\mbox{\tiny h}}+C_0$, which together mean that the phase transition at
$R= R_{\mbox{\tiny h}}$ is inherent to the problem and not just 
an artifact of the achievability scheme, and (ii) a strictly positive
error exponent for every $R < C_0+ R_{\mbox{\tiny h}}$ and zero error--exponent beyond $C_0+\Rh$. 
The reason that this upper bound on the error exponent is somewhat weaker than
the ordinary sphere--packing bound is that here, the auxiliary density,
$g$, is zero--mean and it differs from $f$ only in its variance. The stronger sphere--packing bound, which is also tight in
the range of high rates, is obtained (in ordinary coded communication, without a helper) when the noise under $g$ is allowed to have a mean that is
proportional to the
transmitted signal, e.g., $Z_i\sim g_i=\calN(\theta x_i,\tsigma^2)$, and then, the K--L divergence, $D(g\|f)$, 
is minimized w.r.t.\ both $\theta$ and $\tsigma^2$, and not just the noise variance, $\tsigma^2$,
as is done here. The problem is that here, $x_i$ itself is an arbitrary function of $z^n$, a fact that causes a considerable complication.

\noindent
{\it Proof of Theorem \ref{thm2}.}
Let $Z^n$ be a zero--mean Gaussian vector with covariance matrix $\sigma^2I_n$, $I_n$ being the $n\times n$ identity matrix,
and let $f(z^n)$ denote the corresponding Gaussian density of $z^n$.
Let $g(z^n)$ denote the pdf of an auxiliary Gaussian density with zero mean and covariance matrix $\tsigma^2I_n$.
For a given encoder $\phi$, decoder $\psi$ and helper $T$,
let $\calS_m$, $m=0,1,\ldots,e^{nR}-1$, be the set of noise vectors $\{z^n\}$ for which the decoder errs, that is,
$\calS_m=\{z^n:~\psi(\phi(m,T(z^n))+z^n)\ne m\}$.
Finally, for a given, arbitrarily small $\epsilon > 0$, define
\begin{equation}
        \calA=\left\{z^n:~\sum_{i=1}^n\ln\frac{g(z_i)}{f(z_i)}\le n[D(g\|f)+\epsilon]\right\}.
\end{equation}
Then,
\begin{eqnarray}
        P_{\mbox{\tiny e}}&=&\frac{1}{M}\sum_{m=0}^{M-1}\int_{\calS_m}f(z^n)\mbox{d}z^n\\
        &=&\frac{1}{M}\sum_{m=0}^{M-1}\int_{\calS_m}g(z^n)\cdot\exp\left\{-\sum_{i=1}^n\ln\frac{g(z_i)}{f(z_i)}\right\}\mbox{d}z^n\\
        &\ge&\frac{1}{M}\sum_{m=0}^{M-1}\int_{\calS_m\cap\calA}g(z^n)
        \cdot\exp\left\{-\sum_{i=1}^n\ln\frac{g(z_i)}{f(z_i)}\right\}\mbox{d}z^n\\
        &\ge&\frac{1}{M}\sum_{m=0}^{M-1}\int_{\calS_m\cap\calA}g(z^n)
        \cdot e^{-n[D(g\|f)+\epsilon]}\mbox{d}z^n\\
        &=&e^{-n[D(g\|f)+\epsilon]}\cdot\frac{1}{M}\sum_{m=0}^{M-1}\int_{\calS_m\cap\calA}g(z^n)\mbox{d}z^n\\
        &\ge&e^{-n[D(g\|f)+\epsilon]}\cdot\left[\frac{1}{M}\sum_{m=0}^{M-1}\int_{\calS_m}g(z^n)\mbox{d}z^n-
        \int_{\calA^{\mbox{\tiny c}}}g(z^n)\mbox{d}z^n\right].
\end{eqnarray}
Now, by the weak law of large numbers, the subtracted term, $\int_{\calA^{\mbox{\tiny c}}}g(z^n)\mbox{d}z^n$, tends to zero as $n\to\infty$, for
any $\epsilon > 0$. The first term in the square brackets is the error probability of the same decoder when the noise has
variance $\tsigma^2$. Now, let $\tsigma^2$ be chosen such that
\begin{equation}
        \label{constraint}
        R > \frac{1}{2}\ln\left(1+\frac{P}{\tsigma^2}\right)+R_{\mbox{\tiny h}}.
\end{equation}
Then, according to the converse part of \cite[Theorem 2]{LM20}, the error probability under $g$ is bounded away from zero,
and then the lower bound of the error probability under $f$ is given by an expression of the exponential order of
$$e^{-nD(g\|f)}=\exp\left\{-\frac{n}{2}\left[\frac{\tsigma^2}{\sigma^2}-
\ln\left(\frac{\tsigma^2}{\sigma^2}\right)-1\right]\right\}.$$
The best exponential error bound is clearly achieved by the minimum of $D(g\|f)$ over the set of 
all values of $\tsigma^2$ that comply with (\ref{constraint}).
For $R < \Rh$, this set is empty and hence the minimum is infinity. For $R \ge \Rh+C_0$, $\tsigma^2=\sigma^2$ satisfies (\ref{constraint}), and the
minimum is zero. Finally, in the intermediate range of rates,
(\ref{constraint}), which is
equivalent to
\begin{equation}
	\label{ts}
        \tsigma^2> \frac{P}{e^{2(R-R_{\mbox{\tiny h}})}-1},
\end{equation}
supports any choice of $\tsigma^2$ which is arbitrarily close to the right--hand side of (\ref{ts}), which yields the asserted 
expression of $E_{\mbox{\tiny wsp}}(R)$
in the intermediate range of rates. This completes the proof of Theorem \ref{thm2}. $\Box$

\section{The Modulo--Additive Channel}
\label{ma}

As in \cite{LM20}, here too, we consider also the modulo--additive channel,
\begin{equation}
	Y_i=X_i\oplus Z_i,
\end{equation}
where all three variables take on values in the finite alphabet, $\{0,1,\ldots,K-1\}$, and $\oplus$ designates addition modulo $K$.
In this model, we separate the cases of fixed--rate and variable--rate coding by the helper.

\subsection{Fixed--Rate Coding by the Helper}
\label{fixed}

We begin from the case where the helper employs a fixed--rate code to describe $z^t$. In this case, the best strategy is
to assign $nR_{\mbox{\tiny h}}$ nats to each and every $z^t$ whose probability is not less than $e^{-t\theta}$, 
where $\theta\ge 0$ is chosen as large as possible, but keeping 
the size of the set $\{z^t:~P(z^t)\ge e^{-t\theta}\}$ no larger than $e^{nR_{\mbox{\tiny h}}}$.
This yields
\begin{eqnarray}
	\label{rtheta}
	e^{nR_{\mbox{\tiny h}}}&\ge&\bigg|\left\{z^t:~P(z^t)\ge e^{-t\theta}\right\}\bigg|\\
	&\exe&\exp\left\{t\cdot\max_{\{Q:~-\bE_Q\log P(Z)\le\theta\}}H_Q(Z)\right\}\\
	&\dfn&e^{n\tau r(\theta)},
\end{eqnarray}
and so,
\begin{equation}
	R_{\mbox{\tiny h}}\ge\tau\cdot r(\theta),
\end{equation}
where
\begin{equation}
	\label{rdef}
	r(\theta)=\max_{\{Q:~-\bE_Q\log P(Z)\le\theta\}}H_Q(Z).
\end{equation}
Since $H_Q(Z)$ cannot exceed $\log K$, it is obvious that for any $\tau <  R_{\mbox{\tiny h}}/\log K$, all $t$--vectors
$\{z^t\}$ are represented by this code. The encoder can then fully subtract $Z^t$ (modulo $K$) and thus completely cancel the
noise in the with--help phase and transmit $t\log K=n\tau\log K\approx nR_{\mbox{\tiny h}}$ nats. The error exponent associated
with this phase is therefore strictly infinite. For that purpose, there is no need to let $\tau$ tend to zero. 
However, smaller values of $\tau$ could be helpful in the second phase because it means a longer segment to encode in. 
On the other hand, if $\tau$ is chosen smaller than $R_{\mbox{\tiny h}}/\log K$, the number of error--free nats conveyed, $n\tau\log K$, will
be strictly smaller than $nR_{\mbox{\tiny h}}$. The error exponent
in the second phase would be 
\begin{equation}
(1-\tau)E_{\mbox{\tiny a}}\left(\frac{R-\tau\log K}{1-\tau}\right),
\end{equation}
where, as before, $E_{\mbox{\tiny a}}(\cdot)$ is the error exponent associated with any achievability scheme.
To find the optimal value of $\tau\in(0,1)$ that maximizes this quantity, consider the following chain of equalities:
\begin{eqnarray}
	(1-\tau)E_{\mbox{\tiny r}}\left(\frac{R-\tau\log K}{1-\tau}\right)
	&=&\max_{0\le\rho\le 1}\{(1-\tau)E_0(\rho)-\rho(R-\tau\log K)\}\\
	&=&\max_{0\le\rho\le 1}\{\tau[\rho\log K-E_0(\rho)]+E_0(\rho)-\rho R\}.
\end{eqnarray}
Since $E_0(\rho)/\rho\le\lim_{\rho\to 0}E_0(\rho)/\rho=I(X;Y)\le\log K$, it appears that it is optimal to let
$\tau$ be as large as it may be within the range $[0, R_{\mbox{\tiny h}}/\log K]$, namely $\tau= R_{\mbox{\tiny h}}/\log K$.

An alternative expression of $r(\theta)$, defined as in (\ref{rdef}), is as follows.
\begin{eqnarray}
	r(\theta)&=&\max_{\{Q:~-\bE_Q\log P(Z)\le\theta\}}H_Q(Z)\\
	&=&\max_Q\min_{\lambda\ge 0}[H_Q(Z)+\lambda\{\theta+\bE_Q\log P(Z)\}]\\
	&=&\min_{\lambda\ge 0}\left[\lambda\theta+\max_Q\sum_zQ(z)\log\frac{P^\lambda(z)}{Q(z)}\right]\\
	&=&\min_{\lambda\ge 0}\left[\lambda\theta+\log\left(\sum_zP^{\lambda}(z)\right)\right]\\
	&=&\min_{\lambda\ge 0}[\lambda\theta+(1-\lambda)H_{\lambda}(Z)],
\end{eqnarray}
where $H_\lambda(Z)$ is the R\'enyi entropy of $Z$ of order $\lambda$, defined as
\begin{equation}
	H_{\lambda}(Z)=\frac{1}{1-\lambda}\log\left(\sum_zP^{\lambda}(z)\right).
\end{equation}
The function $r(\theta)$ has the following properties: 
\begin{enumerate}
	\item It is monotonically non--decreasing and
concave. 
	\item For $\theta < \theta_0\dfn\log\frac{1}{\max_zP(z)}$, $r(\theta)=-\infty$ (as the maximization is over an empty set).
	\item For $\theta\ge\theta_\infty\dfn\frac{1}{K}\sum_z\log\frac{1}{P(z)}$, it saturates, that is,
$r(\theta)=\log K$. 
\end{enumerate}

Observe that for a given $\Rh$ and $\tau$, 
there is nothing to gain from selecting a finite value of $\theta$ if it happens to
be larger than $\theta_\infty$, because all $z^t$-sequences are represented anyway, so it is better to enlarge $\theta$ indefinitely
in order to minimize the probability of encoding failure, which is
\begin{equation}
	\mbox{Pr}\left\{P(z^t)< e^{-\theta t}\right\}\exe\exp\left\{-n\tau\min_{\{Q:~-\bE_Q\log P(Z)\ge\theta\}}D(Q\|P)\right\}
	\dfn e^{-n\tau E(\theta)},
\end{equation}
where
\begin{eqnarray}
	E(\theta)&=&\min_{\{Q:~-\bE_Q\log P(Z)\ge\theta\}}D(Q\|P)\\
	&=&\min_Q\sup_{\lambda\ge 0}\left\{\sum_zQ(z)\log\frac{Q(z)}{P(z)}+\lambda\left[\sum_zQ(z)\log P(z)+\theta\right]
	\right\}\\
	&=&\sup_{\lambda\ge 0}\left\{\lambda\theta+\min_Q\sum_zQ(z)\log\frac{Q(z)}{P^{1-\lambda}(z)}\right\}\\
	&=&\sup_{\lambda\ge 0}\left\{\lambda\theta-\log\left[\sum_zP^{1-\lambda}(z)\right]\right\}\\
	&=&\sup_{\lambda\ge 0}\lambda\left[\theta-H_{1-\lambda}(Z)\right].
\end{eqnarray}
The function $E(\theta)$ has the following properties: 
\begin{enumerate}
	\item It is monotonically non--decreasing and convex. 
	\item For
$\theta\le H(Z)$, $E(\theta)=0$. 
		\item For $\theta>\log\frac{1}{\min_zP(z)}$, $E(\theta)=\infty$.
\end{enumerate}
To summarize, the line of thought, in the fixed--rate case considered here, is as follows.
For a given $R_{\mbox{\tiny h}}/\tau=r$, we select $\theta$ according to 
\begin{equation}
	\theta=\theta(r)=\sup_{s\ge 0}[sr+(1-s)H_{1/s}(Z)],
\end{equation}
which is the inverse function of $r(\theta)$,
and then the error exponent associated with the ``with-help'' phase is simply $\tau E(\theta(r))$.
But given $R_{\mbox{\tiny h}}$, we can always select $\tau$ sufficiently small (in particular, $\tau < R_{\mbox{\tiny h}}/\log K$), 
so that $r$ would be as large as desired, and so, the error exponent will be strictly infinite.
Of course, here too, for $R > R_{\mbox{\tiny h}}$, one encodes the extra rate of $R- R_{\mbox{\tiny h}}$ in the other segment,
of length $n(1-\tau)$, without help, as before, and the resulting error exponent is, once again
$$(1-\tau)E_{\mbox{\tiny a}}\left(\frac{R-\tau\log K}{1-\tau}\right),$$
which for $\tau\to 0$, becomes $E_{\mbox{\tiny a}}(R-R_{\mbox{\tiny h}})$.

\subsection{Variable--Rate Coding by the Helper}
\label{var}

For a helper that is allowed to use variable--rate coding, the situation is even simpler than that of fixed--rate coding.
Consider a (universal) data compression scheme where $L(z^t)\approx t\hat{H}_{z^t}(Z)$, where $\hat{H}_{z^t}(Z)$
designates the empirical entropy of $z^t$. This can be accomplished, for example, by a two--part code, whose first part encodes the index
of the type of $z^t$ (using $O(\log t)$ nats) and the second part encodes the index of $z^t$ within the type (using about $t\hat{H}_{z^t}(Z)$ nats).
If this coded information is stored in a finite buffer of length $n\Rh$,
the error in the ``with--help'' phase can only result from
a buffer overflow,
\begin{equation}
	\mbox{Pr}\{L(z^t)\ge nR_{\mbox{\tiny h}}\}\exe\exp\left\{-n\tau\cdot\min_{\{Q:~\tau H(Q)\ge R_{\mbox{\tiny h}}\}}
	D(Q\|P)\right\}.
\end{equation}
As before, if $\tau < R_{\mbox{\tiny h}}/\log K$, the overflow exponent is infinite. By selecting
$\tau$ arbitrarily close to $R_{\mbox{\tiny h}}/\log K$ (from below), we can convey $t\log K=n\tau\log K\approx n R_{\mbox{\tiny h}}$
nats completely error--free. Otherwise, we have an error exponent of
\begin{eqnarray}
	\min_{\{Q:~\tau H(Q)\ge R_{\mbox{\tiny h}}\}}
        D(Q\|P)
	&=&\sup_{\lambda\ge 0}\min_Q\left\{\tau D(Q\|P)+\lambda[R_{\mbox{\tiny h}}-\tau H(Q)]\right\}\\
	&=&\sup_{\lambda\ge 0}\min_Q\left\{\tau\sum_zQ(z)\log\frac{Q^{1+\lambda}(z)}{P(z)}+\lambda R_{\mbox{\tiny h}}\right\}\\
	&=&\sup_{\lambda\ge 0}\min_Q\left\{\tau(1+\lambda)\sum_zQ(z)\log\frac{Q(z)}{P^{1/(1+\lambda)}(z)}+
	\lambda R_{\mbox{\tiny h}}\right\}\\
	&=&\sup_{\lambda\ge 0}\left\{-\tau(1+\lambda)\log\left[\sum_zP^{1/(1+\lambda)}(z)\right]+
	\lambda R_{\mbox{\tiny h}}\right\}\\
	&=&\sup_{\lambda\ge 0}\lambda[R_{\mbox{\tiny h}}-\tau H_{1/(1+\lambda)}(Z)].
\end{eqnarray}
Once again, for $R > R_{\mbox{\tiny h}}$ one encodes the extra rate of $R- R_{\mbox{\tiny h}}$ in the other segment,
of length $n(1-\tau)$, without help, as before.\\

\subsection{The Converse Bound}

The converse bound is obtained using the same ideas as before, except that integrations are replaced by summations.
Owing to \cite[Theorem 8]{LM20}, the WSP bound is given
by the minimum of $D(Q\|P)$ over all noise distributions, $\{Q\}$, for which $R > \min\{I(Q)+R_{\mbox{\tiny h}},\log K\}$,
where $I(Q)=\log K-H(Q)$ is the mutual information induced by a uniformly distributed input and noise governed by $Q$.
For $R < \log K$, this is equivalent to 
\begin{equation}
E_{\mbox{\tiny wsp}}(R)=\min_{\{Q:~I(Q)<R-R_{\mbox{\tiny h}}\}}D(Q\|P),
\end{equation}
which vanishes for 
$R\ge I(P)+R_{\mbox{\tiny h}}\equiv C_0+R_{\mbox{\tiny h}}$, and becomes infinite for $R < R_{\mbox{\tiny h}}$.

\section{The Gaussian Multiple Access Channel}
\label{mac}

Consider now the Gaussian MAC,
\begin{equation}
	Y_i=X_{1,i}+X_{2,i}+Z_i,
\end{equation}
where $X_{1,i}$ and $X_{2,i}$ are the transmitted symbols of the two users at time $i$ and $\{Z_i\}$ is AWGN as before.
The two encoders are subjected to power constraints, $P_1$ and $P_2$, respectively. 
It is well known (see, e.g., \cite[Subsection 15.3.6]{CT06}) that the capacity region of the Gaussian MAC
(without help) is given by
\begin{equation}
	\calC_0=\left\{(R_1,R_2):~R_1\le c(\gamma_1),~R_2\le c(\gamma_2),~R_1+R_2\le c(\gamma_1+\gamma_2)\right\},
\end{equation}
where $\gamma_i=P_i/\sigma^2$, $i=1,2$, and $c(\gamma)=\frac{1}{2}\log(1+\gamma)$.

Now, consider the case, originally studied in \cite{LM20}, where a helper describes the noise by $T_1(z^n)\in\{0,1,\ldots,e^{nR_{\mbox{\tiny h}1}}-1\}$
and $T_2(z^n)\in\{0,1,\ldots,e^{nR_{\mbox{\tiny h}2}}-1\}$ to encoders 1 and 2, respectively, with the limitation $R_{\mbox{\tiny h}1}+
R_{\mbox{\tiny h}2}\le R_{\mbox{\tiny h}}$, where the rate allocation to the two encoders is subjected to optimization.
The transmissions of the two users are then $x_1^n=\phi_1(m_1,T_1(z^n))$ and
$x_2^n=\phi_2(m_2,T_2(z^n))$, where $m_i\in\{0,1,\ldots,e^{nR_i}-1\}$, $i=1,2$, and, as already mentioned,
both transmissions are subject to their corresponding power constraints.

Lapidoth and Marti \cite{LM20} have shown that the capacity region of this configuration is given by the Minkowsky sum of $\calC_0$ and
the triangle formed by the set of rate pairs whose sum does not exceed $R_{\mbox{\tiny h}}$. Equivalently,
it is given by
\begin{equation}
	\calC(R_{\mbox{\tiny h}})=\left\{
	(R_1,R_2):~R_1\le c(\gamma_1)+R_{\mbox{\tiny h}},~R_2\le c(\gamma_2)+R_{\mbox{\tiny h}},~R_1+R_2
	\le c(\gamma_1+\gamma_2)+R_{\mbox{\tiny h}}\right\}.
\end{equation}

From the viewpoint of error exponents, our achievability scheme is slightly different from that of \cite{LM20}.
We can achieve an arbitrarily large error exponent whenever $R_1+R_2< R_{\mbox{\tiny h}}$, as follows. The transmission is divided into
three segments, two of length $t=n\tau$, and one of length $n-2t=n(1-2\tau)$. In the first segment of length $t$, Encoder 2 is silent and
only Encoder 1
transmits. His transmission includes a lattice codeword of dimension $t$ 
minus the quantized noise at rate $R_{\mbox{\tiny h}1}$, exactly as (in, say, the fixed--rate scheme that was described) before. 
In the second $t$-segment, the roles of the two encoders switch: 
Encoder 2 transmits a codeword minus the quantized noise using a rate-$R_{\mbox{\tiny h}2}$ description of the noise, while Encoder 1 
is silent. This scheme supports (essentially) error--free channel coding up to rates 
$R_{\mbox{\tiny h}1}$ and $R_{\mbox{\tiny h}2}$, respectively. If $R_1$ and/or $R_2$
exceed their corresponding help rates, then the excess rates, $\Delta R_i=R_i-R_{\mbox{\tiny h}i}$, $i=1,2$, are conveyed using ordinary
coding for the MAC, without any further help. The error exponent (for small $\tau$) is therefore essentially
$E_{\mbox{\tiny a}}(R_1-R_{\mbox{\tiny h}1},R_2-R_{\mbox{\tiny h}2})=
E_{\mbox{\tiny a}}(R_1-R_{\mbox{\tiny h}1},R_2-R_{\mbox{\tiny h}}+R_{\mbox{\tiny h}1})$, 
where $E_{\mbox{\tiny a}}(\cdot,\cdot)$ is any achievable error exponent for the Gaussian MAC (see, e.g., \cite{Gallager85}, \cite{WAP04}),
and it is interesting to optimize the help--rate allocation
so as to maximize the error exponent. The optimal solution, in this sense, may differ from the optimal solution for achieving the capacity region.
For example, according to \cite{Gallager85}, if $\gamma_1=\gamma_2\dfn\gamma$ and both $R_1$ and $R_2$ are
smaller than $c(\gamma/2)=\gamma/[4(\gamma+2)]$, then the random coding exponent is given by
\begin{equation}
	E_{\mbox{\tiny r}}(R_1,R_2)=\min\left\{c\left(\frac{\gamma}{2}\right)-R_1,
	c\left(\frac{\gamma}{2}\right)-R_2,c(\gamma)-R_1-R_2\right\}
\end{equation}
and so,
\begin{equation}
	E_{\mbox{\tiny r}}(R_1-R_{\mbox{\tiny h}1},R_2-R_{\mbox{\tiny h}2})=
	\min\left\{c\left(\frac{\gamma}{2}\right)-R_1+R_{\mbox{\tiny h}1},
	c\left(\frac{\gamma}{2}\right)-R_2+R_{\mbox{\tiny h}}-R_{\mbox{\tiny h}1},c(\gamma)-R_1-R_2+R_{\mbox{\tiny h}}\right\}
\end{equation}
which is maximized when
\begin{equation}
	c\left(\frac{\gamma}{2}\right)-R_1+R_{\mbox{\tiny h}1}=
	c\left(\frac{\gamma}{2}\right)-R_1+R_{\mbox{\tiny h}}-R_{\mbox{\tiny h}1},
\end{equation}
or, equivalently,
\begin{equation}
	R_{\mbox{\tiny h}1}=\frac{R_1-R_2+R_{\mbox{\tiny h}}}{2};~~
	R_{\mbox{\tiny h}2}=\frac{R_2-R_1+R_{\mbox{\tiny h}}}{2}.
\end{equation}

The WSP upper bound on the error exponent is based on the corresponding converse theorem in \cite{LM20}. It is given by the
minimum divergence between two Gaussian pdfs, $D(\calN(0,\tsigma^2)\|\calN(0,\sigma^2))$, where $\tsigma^2$ is in the set
\begin{equation}
	\label{spset}
	\left\{\tsigma^2:~R_1>c\left(\frac{P_1}{\tsigma^2}\right)+R_{\mbox{\tiny h}}~\mbox{or}~
	R_2>c\left(\frac{P_2}{\tsigma^2}\right)+R_{\mbox{\tiny h}}~\mbox{or}~
	R_1+R_2>c\left(\frac{P_1+P_2}{\tsigma^2}\right)+R_{\mbox{\tiny h}}\right\}
\end{equation}
namely, the minimizing $\tsigma^2$ is given by
\begin{eqnarray}
	\tsigma^2&=&\min\left\{\frac{P_1}{e^{2(R_1-R_{\mbox{\tiny h}})}-1},
	\frac{P_2}{e^{2(R_2-R_{\mbox{\tiny h}})}-1},
	\frac{P_1+P_2}{e^{2(R_1+R_2-R_{\mbox{\tiny h}})}-1}\right\}\nonumber\\
	&=&\sigma^2\cdot\min\left\{\frac{e^{2c(\gamma_1)}-1}{e^{2(R_1-R_{\mbox{\tiny h}})}-1},
	\frac{e^{2c(\gamma_2)}-1}{e^{2(R_2-R_{\mbox{\tiny h}})}-1},
	\frac{e^{2c(\gamma_1+\gamma_2)}-1}{e^{2(R_1+R_2-R_{\mbox{\tiny h}})}-1}\right\},
\end{eqnarray}
which yields
\begin{equation}
	E_{\mbox{\tiny wsp}}(R_1,R_2)=\min\{E_1(R_1),E_2(R_2),E_3(R_1+R_2)\},
\end{equation}
with
\begin{eqnarray}
	E_1(R_1)&=&\frac{1}{2}\left[\frac{e^{2c(\gamma_1)}-1}{e^{2(R_1-R_{\mbox{\tiny h}})}-1}-
	\ln\left(\frac{e^{2c(\gamma_1)}-1}{e^{2(R_1-R_{\mbox{\tiny h}})}-1}\right)-1\right]\\
	E_2(R_2)&=&\frac{1}{2}\left[\frac{e^{2c(\gamma_2)}-1}{e^{2(R_2-R_{\mbox{\tiny h}})}-1}-
	\ln\left(\frac{e^{2c(\gamma_2)}-1}{e^{2(R_2-R_{\mbox{\tiny h}})}-1}\right)-1\right]\\
	E_3(R_1+R_2)&=&\frac{1}{2}\left[\frac{e^{2c(\gamma_1+\gamma_2)}-1}{e^{2(R_1+R_2-R_{\mbox{\tiny h}})}-1}-
	\ln\left(\frac{e^{2c(\gamma_1+\gamma_2)}-1}{e^{2(R_1+R_2-R_{\mbox{\tiny h}})}-1}\right)-1\right].
\end{eqnarray}
$E_{\mbox{\tiny wsp}}(R_1,R_2)$ vanishes if either one of $E_1(R_1)$, $E_2(R_2)$ or $E_3(R_1+R_2)$ vanish, which is the case
when $(R_1,R_2)$ falls outside $\calC(R_{\mbox{\tiny h}})$. On the other hand, $E_{\mbox{\tiny wsp}}(R_1,R_2)=\infty$ whenever
$R_1+R_2< R_{\mbox{\tiny h}}$ (because then the set (\ref{spset}) is empty), 
so here too, the WSP bound is matching the achievability bound at least as far as the qualitative behavior goes.
The same WSP bound was used also in \cite{UKM18}, but in a different context.



\end{document}